\documentclass[a4paper,aps,prb,twocolumn,preprintnumbers,amsmath,amssymb]{
revtex4}
\usepackage{amsmath}
\usepackage{verbatim}
\usepackage{graphicx}

 \newcommand{\ket}[1]{\left|#1\right\rangle} 
 \newcommand{\bra}[1]{\left\langle#1\right|} 

\begin{document}


\title{Spiral spin textures of bosonic Mott insulator with SU(3) spin-orbit coupling}

\author{ Tobias Gra\ss$^1$, Ravindra W. Chhajlany$^{1,2}$, Christine A. Muschik$^{1,3}$,
and Maciej Lewenstein$^{1,4}$}

\address{$^1$ ICFO-Institut de Ci\`encies Fot\`oniques, Av. C.F. Gauss 3, 08860 Barcelona, Spain}
\address{$^2$ Faculty of Physics, Adam Mickiewicz University, Umultowska 85, 
61-614 Pozna{\'n}, Poland}
\address{$^3$  Institute for Quantum Optics and Quantum Information, Austrian Academy of Sciences, 6020 Innsbruck, Austria}
\address{$^4$ ICREA-Instituci\'o Catalana de Recerca i Estudis Avan\c cats, Lluis Campanys 23,
08010 Barcelona, Spain}

\begin{abstract}
We study the Mott phase of three-component bosons, with one particle per site, in an optical lattice by
mapping it onto an SU(3) spin model. In the simplest case  of full SU(3)
symmetry, one obtains a ferromagnetic Heisenberg model. Introducing an
SU(3) analog of spin-orbit coupling, additional spin-spin
interactions are generated. We first consider the scenario of spin-dependent hopping phases, leading to
Dzyaloshinskii-Moriya-type interactions.  They result in the formation of spiral spin textures, which in one dimension can be understood by a local unitary transformation. Applying classical Monte Carlo simulations, we extend our study to two-dimensional systems, and systems with "true" spin-orbit coupling, involving spin-changing hoppings.
\end{abstract}

\maketitle

\section{Introduction}
Quantum many-body physics is to a large extent the physics of interacting spins.
Literally, the spin is an intrinsic property of particles,
formally described by a representation of the SU(2) group. For example, the
fundamental representation of SU(2), expressed in terms of  the Pauli matrices, describes the
1/2-spin, as present, for instance, in electrons. It requires only a small
amount of abstraction to extend the meaning of spin to an arbitrary internal degree of
freedom carried by the particles. Such an extended spin picture is very common in
quantum engineering, where it may denote the different states in which cold
atoms or trapped ions are prepared. While these states may in principal belong
to one hyperfine manifold of the atoms, such that the true atomic spin would
distinguish between the states, this is not necessarily the case. Then the
different states form a so-called \textit{pseudospin} manifold, which, in
contrast to real spin, is independent from the quantum-statistical properties of
the particles \cite{Myatt1997}. In an even wider sense, one might also map external degrees of
freedom onto a spin system. For instance, empty or occupied sites on a lattice
could be seen as a spin-1/2 pointing up or down. This is what occurs in the hard
boson limit of the Bose Hubbard model (cf. Ref.~\onlinecite{mlbook}). The most
common way to derive spin models from Hubbard models is to consider the limit of
strong interactions, and derive an effective Hamiltonian, describing
super-exchange interactions \cite{Anderson1950,Anderson1959}.

\begin {figure}
\centering
\includegraphics[width=0.48\textwidth, angle=0]{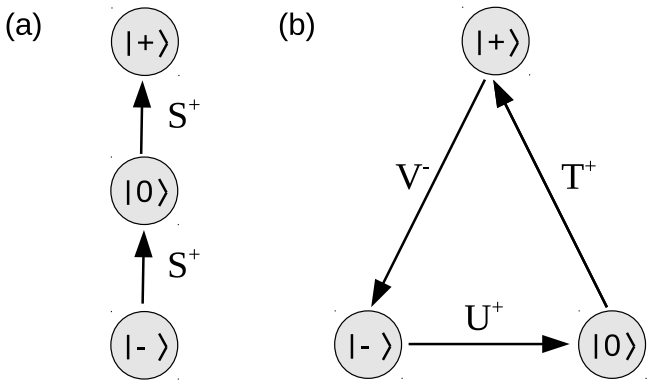}
\caption{\label{Fig1}
{\bf SU(2) vs. SU(3)}: Transitions in a three-level system might span a SU(2)
algebra with a single raising (lowering) operator $S^\pm$, as depicted in (a),
but most generally span a SU(3) algebra with three pairs of raising lowering
operators, $T^{\pm},U^{\pm}$, and $V^{\pm}$.
}
\end {figure}

What is common to these different spin pictures, is its formal aspect: the
operators describing the degree of freedom under scrutiny belong to some
representation of the SU(2) group. In particular, for the case of pseudospin,
one typically has a two-level system. In this case, indeed any local unitary operation
on the spin degree of freedom can be expressed through a single Pauli matrix.
However, if the (pseudo)spin degree of freedom involves more than two states,
the spin picture might become cumbersome. As illustrated in Fig. \ref{Fig1}a for
a three-level system, the transition from state $\ket{+}$ to state
$\ket{-}$ cannot be described in terms of a single SU(2) spin matrix. As SU(2)
has only three generators, there can only be one spin-raising operator, say
$S^+$, which raises state $\ket{-}$ to $\ket{0}$, and $\ket{0}$
to $\ket{+}$, but a direct link from $\ket{-}$ to $\ket{+}$ is missing. To
connect these two states, one has to apply twice the raising operator $S^+$. It
may then be more convenient to turn to a new spin, or better ``isospin'',
picture, in which all operators acting on the degree of freedom span an SU(3)
group. Since this group provides three different raising operators, any unitary
operation on the three-level Hilbert space can then be expressed by a single
SU(3) spin matrix, see Fig. \ref{Fig1}b. This demonstrates that SU(3) spins may
play an important role in systems with three-level constituents ranging from
high-energy systems of elementary particles with their intrinsic three-fold
degrees of freedom like flavor or color \cite{8fold}, to condensed matter
systems with a three-dimensional pseudospin manifold, or quantum optical systems of three level atoms \cite{Shore2001}.
 In these contexts  SU(3) spin models and their ground states have recently been
investigated \cite{greiter2007,toth2010,go13}. Quantum-chaotic behavior in a
SU(3) spin system built of trapped ions has been studied in
Ref.~\onlinecite{su3ions}.

In this paper, we focus on multi-component quantum gases, which have recently attracted a
lot of attention in the wide field of quantum simulation and quantum control
\cite{mlbook,Zhang2014,Scazza2014,Cappellini2014}. In particular, we study those models which arise from a
three-fold pseudospin degree of freedom combined with variants of spin-orbit
coupling. The latter is the important mechanism, which connects external and
internal degrees of freedom of the particles. We stress that  in Nature, spin-orbit coupling exists due to the electric
charges of the particles and the magnetic moments associated with the spins. In
atomic gases, spin-orbit coupling has to be engineered artificially, as has been
pioneered in Ref.~\onlinecite{spielman-sobec} for a Bose-Einstein condensate
with
pseudospin-1/2. Such spin-orbit coupling leads to a rich phase diagram
\cite{galitski,sinha11},  with order-by-disorder induced by quantum fluctuations
\cite{ryan}.
The great versatility of cold atoms allows one to study the effects of spin-orbit
coupling in different physical scenarios, from  Josephson
physics \cite{josephsonSOC} to fractional quantum Hall effect physics \cite{jphysb}.
There are also several proposals to engineer spin-orbit coupling for atoms in an
optical lattice \cite{osterloh,hauke-shakes}, which would affect the Mott
transition \cite{indianpra,iskin2014}, give rise to exotic superfluidity
\cite{indianpra}, and generate a surprisingly rich variety of spin textures in the
Mott phase \cite{trivediBH,galitski12,caiSOC2012,peotta14,alessio}, mostly via
superexchange terms of the Dzyaloshinskii-Moriya type
\cite{dzyaloshinskii,moriya}.

Formally, spin-orbit coupling is a Berry connection, which can be minimally
coupled to the momentum. In the lattice, one makes the Peierls substitution
\[-t \Psi_{i+x}^\dagger \Psi_{i} \rightarrow -t \Psi_{i+x}^\dagger
e^{-i A_x} \Psi_{i}, \]that is, a particle hopping along
$x$-direction (with tunneling amplitude $t$) from site $i$ to site $(i+x)$
experiences a spin rotation described by the gauge potential $A_x$
acting on the spinor $\Psi_i$. In the case of artificial spin-orbit
coupling for pseudospin larger than 1/2, which can be engineered within a
$n$-pod scheme \cite{gediminas-spin1}, this gauge potential may belong to
SU($N$) rather than SU(2). An SU(3) spin-orbit coupling has been
investigated in Ref.~\onlinecite{ryanSU3} in the tight-binding limit, and has
been shown to produce a topologically non-trivial bandstructure.

It is the goal
of this paper to extend the study of SU(3) spin-orbit coupling to the strongly interacting
regime: therefore, in Sec. II,  we derive effective Hamiltonians for
three-component bosons in the Mott phase. After discussing the differences
between spinor and pseudospin gases in the interaction term, we introduce
(pseudo)spin-dependent hopping phases, and arrive at an effective Hamiltonian
with Dzyaloshinskii-Moriya-type interactions. In Sec. III, we solve this
Hamiltonian analytically by means of a local unitary transformation. We find
spiral spin textures, with a periodicity controlled by the hopping phase.
Applying exact diagonalization and classical Monte Carlo simulations, we extend
our study to cases not covered by our analytical solution: chains with periodic
boundary conditions, for which a devil staircase behavior is found, and the case
of two-dimensional lattices.  In Sec. IV, we consider systems with a more
complicated spin-orbit coupling, including spin-changing terms. Finally, in
Sec. V, we comment on the experimental realization of such systems, and
possible ways of detecting the spiral order.

\section{Heisenberg model for a Mott phase of three-component bosons}
Three-level bosons have attracted a lot of attention already since the early days
of Bose-Einstein condensation \cite{ho98,bigelow98,ohmi}, as it occurs quite
frequently that bosonic atoms carry the spin one. Let us therefore start from this most familiar case of spinor gases, and view it from a generalized SU(3) point of view. This prepares for a convenient description of pseudospin gases. Finally, we will turn our attention to SU(3) models with Dzyaloshinskii-Moriya-type interactions, relevant for systems  with (pseudo)spin-dependent hopping.

\subsection{Effective Mott Hamiltonian for spinor bosons}
In spinor gases, the interaction is characterized by two paramters $U_0$ and $U_2$. Discretized on a
lattice, the interaction Hamiltonian reads:
\begin{align}
\label{spinor-int}
 H^{\rm int}_{\rm spinor} = \frac{U_0}{2} \sum_i n_i(n_i-1) + \frac{U_2}{2}
\sum_i ({\bf S}_i^2-2n_i ),
\end{align}
where $n_i$ is the occupation number on site $i$, and ${\bf S}_i$ the (total) spin
operator acting on the particle on site $i$.

The kinetic energy of bosons in a standard optical lattice is well described by
nearest-neighbor tunneling,
\begin{align}
H^{\rm kin}_{\rm standard} = -t \sum_{\langle i,j \rangle} \sum_{\sigma \in
\{+,0,-\}} (a_{i\sigma}^\dagger
a_{j\sigma} + {\rm H.c}),
\end{align}
with $t$ the tunneling amplitude. For $t\ll U_0,U_2$, the system is in the Mott
phase, and insight in its physics can be gained by taking the kinetic part as
a second-order perturbation to the interactions. It has been shown that this
leads to a biquadratic SU(2) Heisenberg model \cite{imambekov,yip}
\begin{align}
\label{biqua}
 H_{SU(2)} = -J \sum_{\langle i,j \rangle} \left[ \cos \theta ({\bf S}_i \cdot
{\bf S}_j) + \sin \theta ({\bf S}_i \cdot {\bf S}_j)^2 \right],
\end{align}
with the parameters $J$ and $\theta$ depending on $t,U_0,$ and $U_2$. This
effective Hamiltonian exhibits spin-nematic and ferromagnetic phases, a
prediction which recently has been confirmed by a full quantum Monte-Carlo
study of the problem \cite{batrouni13}.

As argued in the introduction, the biquadratic SU(2) Hamiltonian (\ref{biqua})
can be rewritten in terms of SU(3) spin matrices, which reduces it to a
quadratic form. While the most general expression is provided in the
appendix, a particularly simple special case is the one of spin-independent
interactions, $U_2=0$, corresponding to $\theta=\pi/4$. The
SU(3) symmetry of the system then leads to an effective SU(3) Heisenberg
Hamiltonian
\begin{align}
\label{heisenberg-su3}
 H_{SU(3)} = - \frac{t^2}{2U_0} \sum_{\langle i,j \rangle}
\sum_{\nu=1}^8
\lambda^{(\nu)}_i \lambda^{(\nu)}_j = - \frac{J}{2\sqrt{2}}
\sum_{\langle i,j \rangle} {\boldsymbol \lambda}_i^T \cdot {\boldsymbol
\lambda}_j,
\end{align}
where we have inserted $J=-\sqrt{2}t^2/U_0$. The $\lambda^{(\nu)}$ are
the eight generators of SU(3), which, for convenience, have been arranged
together to a eight-component vector ${\boldsymbol \lambda} =
(\lambda^{(1)},\dots,\lambda^{(8)})^T$ on the right side of Eq.
(\ref{heisenberg-su3}). The form of Eq. (\ref{heisenberg-su3})
holds for any filling factor if the $\lambda^{(\nu)}$ are chosen a
symmetric SU(3) representation, corresponding to the desired particle
number. An interesting case is the filling with three bosons per site: As
follows from Young tableaux calculus, the Hilbert space of three qutrits
consists of a fully symmetric decuplet, an fully antisymmetric singlet, and two
octets of mixed symmetry, ${\bf 3} \otimes {\bf 3} \otimes {\bf 3} = {\bf 10}
\oplus {\bf 8} \oplus {\bf 8} \oplus {\bf 1}$. According to the bosonic nature
of the atoms, the low-energy Hilbert space is given by the {\bf
10} representation of SU(3), which is implemented in Nature as the baryon
decuplet.

In the following, however, we restrict ourselves to the simpler case of one
boson per site. The low-energy Hilbert space then is three-dimensional, and one
works in the fundamental representation of SU(3). Accordingly, we associate the
eight Gell-Mann matrices with the $\lambda^{(\nu)}$, cf.
Ref.~\onlinecite{8fold}.
Note that six of these matrices are related to the raising/lowering operators
of Fig. \ref{Fig1}b: $(\lambda^{(1)} \pm i \lambda^{(2)}) = 2 T^\pm, \
(\lambda^{(4)} \pm i \lambda^{(5)}) = 2 V^\pm, \ {\rm and} \ (\lambda^{(6)} \pm
i \lambda^{(7)}) = 2 U^\pm$. The other two matrices are diagonal:
$\lambda^{(3)} = {\rm diag}(1,-1,0), \ {\rm and} \ \lambda^{(8)} =
\frac{1}{\sqrt{3}}{\rm diag}(1,1,-2)$.

\subsection{Effective Mott Hamiltonian for bosons with SU(3) pseudospin}

The biquadratic form of Eq. (\ref{biqua}) holds due to the invariance of
$H=H^{\rm kin}_{\rm standard} + H^{\rm int}_{\rm spinor}$ under spin rotations.
If, instead of considering Bose gases with an intrinsic spin, one switches to
pseudospin-1 systems, that is, a trinary mixture of bosons, this symmetry is
likely to be broken. In particular, one might not expect interactions as
described by Eq. (\ref{spinor-int}). Neglecting spin-changing collisions, the
most general two-body contact interaction would be given in terms of six
interaction parameters $U_{\sigma\sigma'}$:
\begin{align}
\label{pseudo}
 H^{\rm int}_{\rm pseudospin} = U_{\sigma\sigma'} \sum_i \ket{\sigma\sigma'}_i
\bra{\sigma\sigma'}_i,
\end{align}
for pairs of particles in state $\ket{\sigma}$ and $\ket{\sigma'}$
interacting on site $i$.  The effective Mott Hamiltonian to $H=H^{\rm kin}_{\rm
standard} + H^{\rm int}_{\rm pseudospin}$ is of the form
\begin{align}
\label{effpseudo}
H_{\rm eff} = &- \sum_{\langle i,j
\rangle} \Bigg[ \sum_{\nu=1}^8 J_\nu \lambda^{(\nu)}_i
\lambda^{(\nu)}_j +
J_{38}
(\lambda^{(3)}_i \lambda^{(8)}_j +
\nonumber \\ &
\lambda^{(8)}_i \lambda^{(3)}_j) \Bigg] + \sum_i \left[h_3 \lambda_i^{(3)}
+ h_8 \lambda_i^{(8)} \right].
\end{align}
Now the interaction parameters $J_\nu$  are spin-dependent, there is an
additional interaction $J_{38}$, and magnetic-field-like terms $h_3$ and $h_8$.
All parameters are given in appendix \ref{app-pseudo}.

\subsection{Effective Mott Hamiltonian in the case of spin-dependent tunneling phases and generalized Dzyaloshinskii-Moriya interactions}

We now turn our attention to systems where the standard hopping is replaced by a spin-orbit coupled kinetic term. Recently, spin-orbit coupling in two-component Bose gases has been shown to
give rise to effective Dzyaloshinskii-Moriya
interactions in the low energy description of the Mott insulator phase \cite{trivediBH,galitski12,caiSOC2012,peotta14}, that is an
antisymmetric spin-spin interaction of the form ${\bf D}\cdot ({\bf S}_i \times
{\bf S}_j)$, characterized by the Dzyaloshinskii-Moriya vector ${\bf D}$.
The presence of such interaction leads to a variety of different spin
textures in the ground state.

In this section, we derive an effective Hamiltonian for
three-component bosons, which exhibits Dzyaloshinskii-Moriya type interactions
generalized to SU(3) spins.  For this, it is sufficient to consider
pseudospin-dependent tunneling phases. If these phases were spatially
dependent, this would mimic a magnetic field acting differently on the three
components, corresponding to different electric charges of the pseudospin
states. Note that such a situation occurs naturally in the quark model, where
the up quark has an electric charge of $2/3e$, while down quark and strange
quark have a charge of $-e/3$.

We start our analysis by writing down the general form of the kinetic term in
the presence of a gauge potential,
\begin{align}
 H^{\rm kin}_{\rm SOC} = -t \sum_i \sum_{{\bf r} \in \{\hat e_x,\hat e_y\}}
{\bf a}_{i+{\bf r}}^\dagger e^{-i {\bf A}\cdot{\bf r}} {\bf a}_i + {\rm H.c.},
\end{align}
where we have introduced a vector notation for the three bosonic components ${\bf a}_i^\dagger =
(a_{i+}^\dagger,a_{i0}^\dagger,a_{i-}^\dagger)$. Spin-orbit coupling is obtained by choosing
the gauge potential to be sensitive to the (psedo)spin.  We have assumed a two-dimensional square lattice in the notation above.
A one-dimensional system is obtained by freezing out the hopping between adjacent one-dimensional chains.

Except in Sec. IV,  we shall consider the simplest choice for the gauge potential corresponding to diagonal $A_x$ and $A_y$.
In the generic 2-dimensional case we choose standard gauge-free hopping in one direction, say $y$, {\it i.e.} $A_y={\bf 1}$.  The non-trivial gauge choice is made in the x-direction (also chosen as the chain direction for 1-d lattices) with $A_x={\rm
diag}(\alpha,\beta,\gamma)$. In the effective Mott Hamiltonian, this will lead
to modified spin interactions along the $x$-direction, as the
superexchange terms gain a phase factor.

Let us, in the first place, assume the important special case of SU(3)-symmetric
interactions of strength $U$. The hopping in $x$-direction then yields the effective
Hamiltonian:
\begin{align}
\label{hx}
 h_x =& -\frac{t^2}{2U} \sum_i \Bigg[ \sum_{\nu=1}^8 J_{\nu}
\lambda_i^{(\nu)} \lambda_{i+x}^{(\nu)}
+
\nonumber \\ &
\sum_{\nu,\mu} V_{\nu\mu} \left(\lambda_i^{(\nu)}
\lambda_{i+x}^{(\mu)} -\lambda_i^{(\mu)}
\lambda_{i+x}^{(\nu)} \right)\Bigg],
\end{align}
with
\begin{align}
\label{JV}
& J_{1} = J_{2} =\cos(\alpha-\beta),  \  J_{4}  = J_{5} =\cos(\alpha-\gamma), \nonumber \\ &
J_6 = J_7 =\cos(\beta-\gamma), \  J_3 = J_8 =1, \\ \nonumber
&
V_{12} = \sin(\alpha-\beta), \  V_{45} = \sin(\alpha-\gamma), \ V_{67}
= \sin(\beta-\gamma).
\end{align}
All other $V_{\nu\mu}$ are zero. Apparently, the second term in Eq.
(\ref{hx}) describes an antisymmetric spin-spin interaction. To make the
analogy to the Dzyaloshinskii-Moriya interaction as close as possible, we
introduce a ``vector product'' of eight-dimensional SU(3) vectors:
\begin{align}
 {\bf u} \times {\bf v} \equiv f^{ijk} u_i v_j {\bf \hat e}_k,
\end{align}
with $f^{ijk}$ the antisymmetric structure constants of SU(3), cf.
Ref.~\onlinecite{8fold}. If $ \sin(\alpha-\beta) = \sin(\alpha-\gamma) -
\sin(\beta-\gamma)$, the effective Hamiltonian $h_x$ can be written as
\begin{gather}
\label{hxDM}
 h_x =   -\frac{t^2}{2U} \sum_i \left[ \sum_{\nu=1}^8 J_{\nu}
\lambda_i^{(\nu)} \lambda_{i+x}^{(\nu)}
+ {\bf  D} \cdot ({\boldsymbol \lambda}_i \times {\boldsymbol
\lambda}_{i+x} )\right],
\end{gather}
with the $J$'s given in Eq. (\ref{JV}), and the non-zero entries of the
Dzyaloshinskii-Moriya vector given by
\begin{align}
 D_{3} &= \sin(\alpha-\beta), \ \  D_{8} = \frac{1}{\sqrt{3}}[
\sin(\alpha-\gamma) + \sin(\beta-\gamma)].
\end{align}
In the discussion below, we consider Hamiltonian's of the type given in Eq.~(\ref{hxDM}) determined by the parameter $\alpha$, with the choice  $\gamma=\alpha, \beta=0$.

As will be discussed in greater detail below, the Dzyaloshinskii-Moriya interaction does not fully lift the huge degeneracy of the  ferromagnetic SU(3) Heisenberg model. One way to obtain a unique ground state is to break the SU(3) symmetry of the interactions. For instance, by strengthening interactions between identical particles, $U$, relative to interactions between particles of different (pseudo)spin, $U_{\rm inter}$,  one might force the system into a fully unpolarized state, with equal occupation numbers in all spin states, $N_+=N_0=N_-$. The effective spin Hamiltonian resulting from these interactions is given again by Eq. (\ref{hx}), but now with the parameters
\begin{align}
\label{hmu}
&
 J_1 =J_2 = \frac{1}{\mu} \cos \alpha,  \  J_4 =J_5=
\frac{1}{\mu}, \nonumber \\ &  J_6=J_7 = \frac{1}{\mu} \cos \alpha, \  J_3 =J_8 =
\frac{2\mu-1}{\mu}, \\ \nonumber
&
V_{12} = \frac{1}{\mu}\sin \alpha, \  V_{45} = 0, \ V_{67}
= -\frac{1}{\mu}\sin \alpha,
\end{align}
where $\mu \equiv U_{\rm inter}/U$ is the ratio between the two
interaction strengths.

\section{Spiral order phases}
For simplicity, we consider first spin models stemming from SU(3) symmetric
interactions, Eq. (\ref{hx}) with parameters from Eq. (\ref{JV}), for discussing
the effect of  the Dzyaloshinskii-Moriya term. We will separately study
1D systems with open and closed boundary, as well as 2D systems with closed
boundary.

\subsection{ Effect of the Dzyaloshinskii-Moriya term in 1D with open boundary}

Freezing the dynamics in the $y$-direction, it is possible  to solve the problem analytically under open boundary conditions.
It is convenient to introduce a spin rotation matrix
$U_i[\alpha]$:
\begin{align}U_i[\alpha]\equiv
 \left(
\begin{array}{ccc}
e^{-i \alpha} & 0 & 0 \\
0 & 1 & 0 \\
0 & 0 & e^{-i \alpha}
\end{array}
\right),
\end{align}
acting on site $i$. For a chain of $N$ spins, labeled $1,\dots,N$ we
introduce
\begin{align}
\label{unitary}
 {\bf U} \equiv \otimes_{k=1}^{N} U_k[(N-k)\alpha],
\end{align}
that is, each pair of neighboring spins is rotated relative to each other by an
angle
$\alpha$. The Hamiltonian $h_x$ from Eq. (\ref{hx}) can then be written
as
\begin{align}
 h_x = -\frac{t^2}{2U} {\bf U}^\dagger \left( \sum_{i=0}^{N-2}
{\boldsymbol \lambda}_i^T {\boldsymbol \lambda}_{i+1}  \right) {\bf U}.
\end{align}
This means that $h_x$ can be mapped onto the ferromagnetic SU(3) Heisenberg
model via a unitary transformation, and both Hamiltonians have the same
energy spectrum. In particular, the ground state degeneracy of the
ferromagnetic Heisenberg Hamiltonian is not lifted by the Dzyaloshinskii-Moriya
term.

However this solution allows one to gain insight into the magnetic correlations in the SU(3) system by mapping the ground states of ferromagnetic model to the Dzyaloshinskii-Moriya Hamiltonian.
The ground state manifold of the SU(3) symmetric Heisenberg ferromagnet corresponds to the maximal symmetric representation of SU(3) on N sites and is therefore spanned by the
symmetric Dicke states, {\it i.e.}, the symmetric superposition of states with
$N_+$ particles in $\ket{+}$, $N_0$ particles in $\ket{0}$, and
$N_-$ particles in $\ket{-}$, where $N_+ +N_0+N_-=N$. Therefore, the number
of degenerate ground states is given by $d=\sum_{N_+=0}^N \sum_{N_0=0}^{N-N_+}
= \frac{1}{2} (N+1)(N+2)$. In the presence of the Dzyaloshinskii-Moriya
interactions, each Fock state contributing to an eigenstate gains a phase
through the unitary transformation, Eq. (\ref{unitary}).

To illustrate this phase acquirement, let us first consider a chain of $N=3$ spins. This yields $d=10$ ground states. Three of them are given in terms of a single Fock state, namely
the states $\ket{+++}
\equiv \ket{+}_{i=1}\otimes\ket{+}_{i=2}\otimes\ket{+}_{i=3}$, $\ket{000}$, and
$\ket{---}$. These states are, apart from a irrelevant overall phase, invariant
under $\bf U$. The remaining seven ground states are transformed by the
Dzyaloshinskii-Moriya interactions. Let us, as an important example, pick out
the fully unpolarized state with one $+$ particle, one $-$ particle, and one $0$ particle.
This is the unique ground state of the isotropic ferromagnetic SU(3) model in a subspace of the Hilbert space with $N_+=N_-=N_0$. Note that these numbers are constants of  motion.

\begin{widetext}
For $N=3$, the fully unpolarized Dicke state reads
\begin{align}
\Psi_{+0-} \equiv \frac{1}{\sqrt{6}} (\ket{+0-} + \ket{+-0} + \ket{0+-}+\ket{0-+} + \ket{-+0} + \ket{-0+}).
\end{align}
The ground state of $h_x$ in the fully unpolarized sector reads
\begin{align}
\tilde \Psi_{+0-} = {\bf U}^\dagger \Psi_{+0-} = \frac{1}{\sqrt{6}}
(e^{2 i\alpha} \ket{+0-} + e^{3 i\alpha} \ket{+-0} + e^{ i\alpha} \ket{0+-} + e^{i\alpha} \ket{0-+} + e^{3 i\alpha} \ket{-+0} + e^{2i\alpha} \ket{-0+}).
\end{align}
\end{widetext}
This is easily generalized to any $N$ being a multiple of 3. 
We note here that breaking the SU(3) symmetry of the interactions with $\mu<1$, 
as in Eq.(\ref{hmu}) (and $\alpha=0$), yields a unique ground state  originating
from the unpolarized Dicke state. The overlap of this ground state with the
unpolarized Dicke state is large for the system sizes for which we have
performed exact diagonalization, e.g for $\mu=0.75, N=9$, it is approximately
0.95.  The effect of the phases $\alpha \neq 0$ can again be taken into account
by the transformation Eq.(\ref{unitary}), which at any $\alpha$ produces the
correct ground state with a fidelity $>0.99$ for $\mu=0.75, N=9$. Therefore the
discussion below holds to a good approximation also for the case when the
Dzyaloshynskii-Moriya type Hamiltonian ground state is unique.

It can easily be seen that due to strong entanglement in this state, the single site density matrices are all propotional to the identity and therefore the spin (traceless Gellmann matrices) averages to zero, $\langle \lambda^{(i)} \rangle=0$, for arbitrary $N$. By combinatoric analysis of the Dicke state, the spin-spin correlations can be evaluated and they show an interesting behavior as a function of $\alpha$:
\begin{align}
\label{qcorr}
\langle
\lambda_i^{(1)} \lambda_{i+r}^{(1)} + \lambda_i^{(2)} \lambda_{i+r}^{(2)}
 \rangle &=
\frac{4N}{9(N-1)}\cos (r\alpha),
\\ \nonumber
\langle
\lambda_i^{(4)} \lambda_{i+r}^{(4)} + \lambda_i^{(5)} \lambda_{i+r}^{(5)}
 \rangle &=
\frac{4N}{9(N-1)},
\\ \nonumber
\langle
\lambda_i^{(6)} \lambda_{i+r}^{(6)} + \lambda_i^{(7)} \lambda_{i+r}^{(7)}
 \rangle &=
\frac{4N}{9(N-1)}\cos (r\alpha),
\\ \nonumber
\langle
\lambda_i^{(3)} \lambda_{i+r}^{(3)}
\rangle &=0, \\ \nonumber
\langle
\lambda_i^{(8)} \lambda_{i+r}^{(8)}
 \rangle &= 0.
\end{align}
The spin structure can thus be viewed as spirals in the $\lambda^{(1)}-\lambda^{(2)}$- and the $\lambda^{(6)}-\lambda^{(7)}$-plane, and a ferromagnet in the $\lambda^{(4)}-\lambda^{(5)}$-plane.

We note that spiral order has also been found in SU(2) systems with spin-orbit coupling \cite{trivediBH,galitski12,caiSOC2012}, but the coexistence of ferromagnetic and spiral order is a feature only found in the SU(3) system.
Moreover, the phase angle $\alpha$ can be used to freely tune the periodicity of
the spirals. In contrast to the systems in
Refs.~\onlinecite{trivediBH,galitski12,caiSOC2012}, we have not assumed
off-diagonal spin-orbit coupling, easing the experimental realization of the
system studied here.

\subsection{Effect of the Dzyaloshinskii-Moriya term with periodic boundary conditions}

The unitary transformation that gauges away the Dzyaloshinskii-Moriya term in the Hamiltonian in the previous section fails, in general, to achieve this completely in the case of a periodically closed boundary. Indeed, the transformation gauges away the  Dzyaloshinskii-Moriya term in the entire bulk of the lattice leaving only an uncompensated, frustrated  boundary term.
The reason for this is the phase relation between the first and the last spin demanded by the periodic boundary condition. Following the logic of the previous section, fixing the phase transformations on the first and last spins results in a phase mismatch of $N \alpha$ on the bond connecting these spins.

As such the magnetic ordering is therefore expected to be qualitatively very similar to that described in the last section.
This is exactly true when the unitary transformation has no conflict with the
periodic boundary as in the special cases when $0 = {\rm mod}[N \alpha, 2\pi]$,
{\it i.e.}, for $\alpha$ being an integer multiple of $2\pi/N$. Interestingly,
our results from exact diagonalization suggest that the ground state
is well described by a unitary transformation for \textit{any} Hamiltonian
parameter $\alpha$. Certainly, the input parameter to the unitary transformation
have to be chosen from the discrete set $\tilde\alpha= m 2\pi/N$. The integer
$m$ is a kind of a ``winding number'' counting the number of times a spin spiral
wraps around the axis throughout the system. This transformation parameter,
$\tilde \alpha$, depends on the Hamiltonian parameter, $\alpha$, through a devil
staircase-like function:
\begin{align}\label{staircase}
 m=j \ \ {\rm for \ } j \in {\cal Z}: \ \frac{2\pi}{N} (j-\frac{1}{2}) \leq
\alpha < \frac{2\pi}{N} (j+\frac{1}{2}).
\end{align}
We emphasize here, that this unitary transformation whilst providing a description of the ground state, for arbitrary $\alpha$, does not unitarily connect between the Heisenberg Hamiltonian and the Dzyaloshinskii-Moryia-type Hamiltonians. This reflects in an energy spectrum depending on $\alpha$.

Due to the discreteness of the input parameters to the unitary transformation,
the finite system with periodic boundary conditions exhibits clearly separated
phases, characterized by the winding number $m$. From this, one obtains the
periodicity of the spin pattern, expressed in the size of the unit cell $P$, as
\begin{align}
\label{unitcell}
 P= \frac{L}{m}p,
\end{align}
where $L$ is the length of the system, and $p$ is the smallest integer such
taht $P$ is integer. If $m$ and $L$ are coprime, the system is incommensurate,
and we have $P=L$. It is worth to notice that for $m=N/2$, that is in the
vicinity of $\alpha=\pi/2$, the unit cell size is 2, that is, the spiral order
becomes antiferromagnetic.

\subsection{ Effect of the Dzyaloshinskii-Moriya term in 2D}

\begin{figure*}
\centering
\includegraphics[width=0.98\textwidth, angle=0]{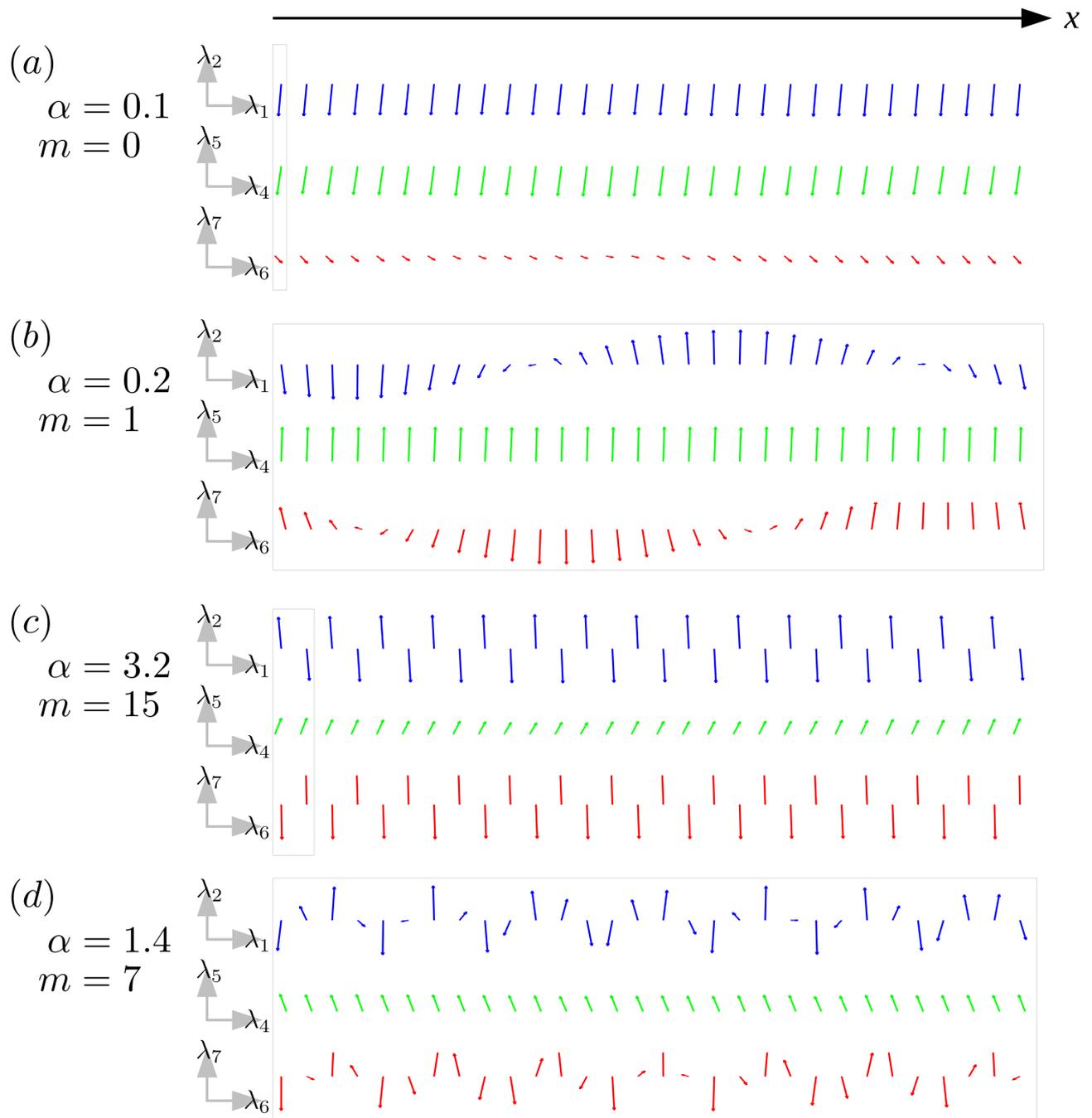}
\caption{\label{Fig2} (Color online)
{\bf Spin textures in the presence of spin-dependent tunneling phases}:
Simulating annealing results for a $30\times 30$ lattice with periodic
boundary for different tunneling phases $\alpha$. We plot projections
of the local SU(3) spin vectors ${\boldsymbol \lambda}$ into different spin
planes. The $\lambda^{(3)}-$ and $\lambda^{(8)}-$component is not plotted, as
it is zero. Since the system behaves ferromagnetic along $y$-direction, we only
show the behavior along $x$.
In (a), despite non-zero $\alpha$, the system remains ferromagnetic in all spin
components. In (b--d), only the
$\lambda^{(4)}-$ and $\lambda^{(5)}-$component behaves ferromagnetic, while the
other spin components show spiral order. In (b), we plot the case with only one
spiral, while in (c) we plot the case with maximum winding number, that is,
$N/2$ spirals, or antiferromagnetic order. In (d), we give an example for an
incommensurate winding number, $m=7$. For all cases, we plot the corresponding
unit cell.
}
\end{figure*}

\begin{figure*}
\centering
\includegraphics[width=0.98\textwidth, angle=0]{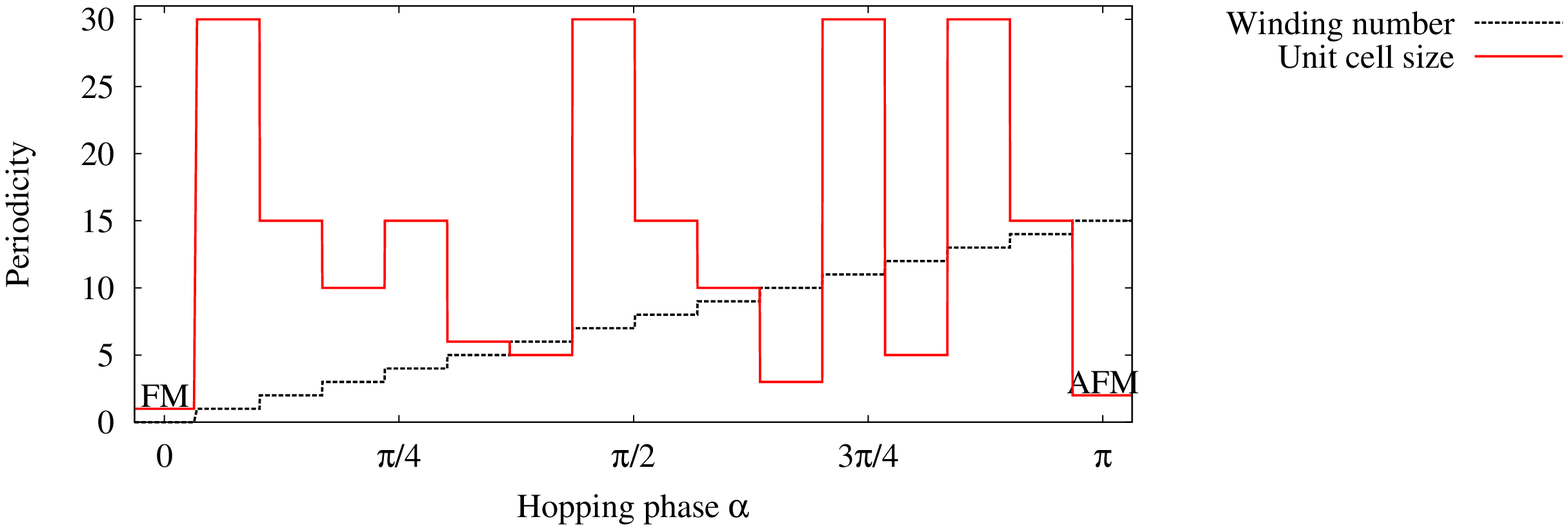}
\caption{\label{Fig3} (Color online)
{\bf System periodicity in the presence of a spin-dependent tunneling phase
$\alpha$:} We consider a system with 30 sites along the direction of
non-trivial hopping, and plot the winding number of the spirals and the
corresponding size of the unit cell as a function of $\alpha$. The figure
represents our MonteCarlo results for a two-dimensional system, cf. \ref{Fig2},
but also illustrates the essence of the staircase formula, Eq.
(\ref{staircase}).
}
\end{figure*}

We now turn to  a two-dimensional system, under periodic boundary conditions, with spin-orbit coupling in one direction, and standard hopping in the other direction. Again, we are not able to map the problem
onto the Heisenberg model by means of the unitary transformation of Eq.
(\ref{unitary}).

We obtain here the classical ground state phase of the 2-D model. In order to achieve this,  the
8-dimensional SU(3) spins are treated classically by replacing them with their expectation values and finding the classical spin configuration on the lattice minimizing the energy.
 The on-site vector $\vec{\lambda} = \frac{\sqrt{3}}{2}\langle {\boldsymbol \lambda} \rangle$ can be seen as a generalized Bloch vector, where the prefactor is needed to attain unit normalization.
It is important to note a crucial
difference w.r.t the well-known SU(2) case: A pure state of a two-level system is
characterized by two angles, a mixing angle, and a relative phase, which
perfectly maps onto a sphere. A pure state in a three-level system is, up to a
irrelevant overall phase, of the form:
\begin{align}
\label{qtrit}
 \ket{\Psi} = e^{i \xi_1} \sin\theta \cos\phi \ket{+} + e^{i \xi_2} \sin\theta
\sin\phi \ket{0} + \cos\theta \ket{-},
\end{align}
with $0\leq\theta,\phi<\frac{\pi}{2}$, and $0\leq\xi_1,\xi_2<2\pi$. This means
that the state is parametrized by only four real numbers, and the
eight-component vector $\vec{\lambda}$ has to be restricted to the
corresponding four-dimensional subspace.

In the classical limit, the spin-spin Hamiltonian provides an
energy functional which is numerically minimized by simulated annealing. The
latter is implemented by starting from an arbitrary initial configuration, which
is locally updated through a Metropolis algorithm. We have repeatedly carried
out the simulated annealing for a lattice with $30\times30$ spins.

For the case of SU(3)-symmetric interaction, Eq. (\ref{JV}), the simulated annealing
always led to a ferromagnet fully polarized within the $\lambda^{(3)}-\lambda^{(8)}$
plane. This can be understood by noting that on the classical level the
Dzyaloshinskii-Moriya interactions do not affect the part of the system which
is polarized in the $\lambda^{(3)}-\lambda^{(8)}$ plane, in which the energy per spin
may remains minimal by forming a ferromagnet.

To force the system into a less trivial configuration than the ferromagnetic
one, we have to make the polarization in the $\lambda^{(3)}-\lambda^{(8)}$ plane
less favorable. This can again be achieved by tuning pseudospin-dependent interactions away from the SU(3)-symmetry point, as described in Eq. (\ref{hmu}). For $\mu<1$, polarization within the
$\lambda^{(3)}-\lambda^{(8)}$ plane becomes energetically less favorable, and the
effect of the Dzyaloshinskii-Moriya interactions is expected to be enhanced.

Accordingly, our simulated annealing  lattice with pseudospin interactions
characterized by $\mu=0.75$ yields interesting spin patterns: As illustrated in
Fig. \ref{Fig2}, there is spiral order along the $x$-axis, and ferromagnetic
order along the $y$-axis with the standard hopping.
The spiral ordering occurs only in the 1-,2-,6-, and
7-components of $\vec{\lambda}$, that is, in those components directly affected by the
Dzyaloshinskii-Moriya term.
As expected, the 3- and 8-components of $\vec{\lambda}$ are
vanishingly small, and the 4- and 5-components of
$\vec{\lambda}$ , unaffected by the Dzyaloshinskii-Moriya interaction, show ferromagnetic order.

The periodicity of these textures along the $x$-axis depends on the hopping
phase $\alpha$, in exactly the same way as in the 1D system with periodic
boundary conditions, cf. Eqs. (\ref{staircase}-\ref{unitcell}). We illustrate
this by plotting the winding number of the spirals and the size of the unit cell
in Fig. \ref{Fig3} for a $30\times30$ lattice. Indeed, the spiral order found
along $x$ is precisely the classical version of the quantum solution we obtained
in 1D. To see this, we evaluate spin-spin correlations in the classical
solution. The site-dependent parametrization of $\vec{\lambda}$ that is in 
agreement with the results of simulated annealing, is given by:
\begin{align}
\vec{\lambda}_{x,y}= \frac{1}{\sqrt{3}} \left(  \cos(x \tilde\alpha),
\sin(x \tilde\alpha), 0,1,0, \cos(x \tilde\alpha),
-\sin(x \tilde\alpha)
\right),
\end{align}
where $\tilde\alpha$ is an integer multiple of $2\pi/L_x$, with $L_x$ the size of the lattice along the $x$-direction. The value of  $\tilde\alpha$ depends on $\alpha$ according to the devil staircase function ({\ref{staircase}) described in the previous section. Using this, the correlations are:
 \begin{align}
\label{ccorr}
\lambda_i^{(1)} \lambda_{i+r}^{(1)} + \lambda_i^{(2)} \lambda_{i+r}^{(2)}
 &=
\frac{1}{3}\cos (r\alpha),
\\ \nonumber
\lambda_i^{(4)} \lambda_{i+r}^{(4)} + \lambda_i^{(5)} \lambda_{i+r}^{(5)}
 &=
\frac{1}{3},
\\ \nonumber
\lambda_i^{(6)} \lambda_{i+r}^{(6)} + \lambda_i^{(7)} \lambda_{i+r}^{(7)}
 &=
\frac{1}{3}\cos (r\alpha),
\\ \nonumber
\lambda_i^{(3)} \lambda_{i+r}^{(3)}
 &=0, \\ \nonumber
\lambda_i^{(8)} \lambda_{i+r}^{(8)}
 &= 0.
\end{align}
Taking into account the normalization constant in the definition of
$\vec{\lambda}$, the spin-spin correlators in the classical configuration are
identical to the quantum ones of Eq. (\ref{qcorr}) in the limit of large number
of spins $N$.

\section{Off-diagonal spin-orbit coupling}

\begin{figure}
\centering
\includegraphics[width=0.46\textwidth, angle=0]{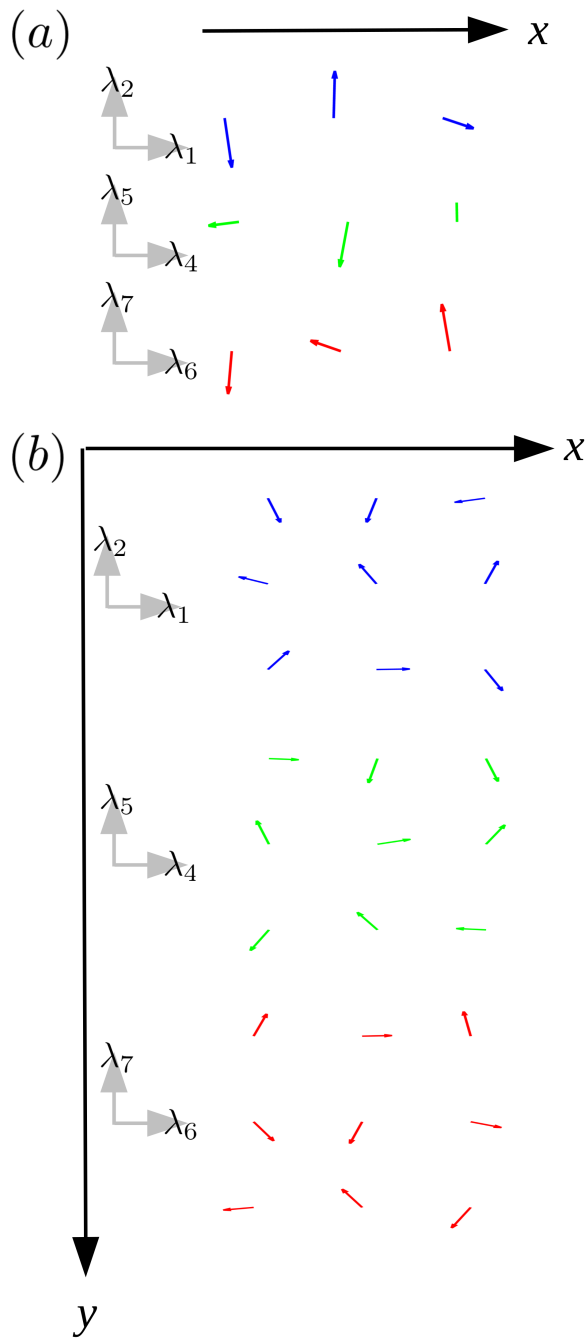}
\caption{\label{Fig4}
(Color online)
{\bf Spin textures in the presence of off-diagonal spin-orbit coupling}: The
plots show the projection of the local SU(3) spin vectors ${\boldsymbol
\lambda}$ into different spin planes, for a Mott system with the spin-orbit
coupling of Eq. (\ref{SOC}). We have studied a lattice of 30 sites in each
direction, but here we only plot the unit cell. In (a), we restrict the dynamics
to the $x$-direction, while (b) shows the results for the two-dimensional
system. 
}
\end{figure}

In the previous section, we discussed the simplest spin-orbit coupling,
consisting only of spin-dependent tunneling phases or abelian spin-orbit coupling. We found that this can give rise to interesting pseudospin textures, but it is not sufficient to fully remove the huge ground state degeneracy of the Heisenberg model. The latter was achieved by breaking the SU(3)-symmetry in the interactions, e.g. by making atoms in the same pseudospin state more repulsive that atoms in different pseudospin states. Furthermore, even in this case, the spiral structures could be traced back to   effects of a gauge transformation acting on the (anisotropic) Heisenberg SU(3) model.

Furthermore, the precise interaction Hamiltonian for spin-orbit coupled quantum
gases is a delicate issue. One has to bear in mind that schemes to
generate the spin-orbit coupling typically dress the atoms. Such dressing might
have significant consequences for the interactions, cf.
Refs.~\onlinecite{williams12,bruno-natcomm}. It seems, however, to be a
reasonable
assumption to consider spin-independent interactions. Nature provides atomic
manifolds with (almost) SU($N$)-symmetric interactions, and if the dressed
states are composed only from states of such manifolds, interactions should
remain spin-independent.

In this section, we therefore analyze a more complicated (non- abelian) spin-orbit coupling, which
includes spin-changing hopping processes  with the assumption of  SU(3) symmetric interactions. As we shall see, such terms again lead to non-trivial
spin textures, but they also remove the degeneracies of the Heisenberg model.

A truly SU(3) spin-orbit coupling has been discussed by Barnett, Boyd, and
Galitski in Ref.~\onlinecite{ryanSU3}. It is described by the vector potential
\begin{align}
\label{SOC}
 (A_x,A_y) = \frac{\pi}{3}\left( -\frac{2}{\sqrt{3}} (\lambda^{(2)} - \lambda^{(5)} +
\lambda^{(7)} ), \lambda^{(3)} + \sqrt{3} \lambda^{(8)} \right).
\end{align}
With this choice, motivated by a mapping onto the SU($N$) Hofstadter model, the
non-interacting system has been shown to exhibit a topological non-trivial
bandstructure.

We now derive the effective Hamiltonian in the strongly interacting limit corresponding to the spin-orbit coupling in Eq.(\ref{SOC}).
We first note that along the $y$-direction the spin-orbit
coupling remains diagonal, and our analysis of the previous section applies.
The interesting, off-diagonal feature of the spin-orbit coupling affects the
hopping along the $x$-direction. Explicitly, it reads
\begin{align}
 H^{\rm kin}_{{\rm SOC}, x} = -t \sum_i {\bf a}_{i+1}^{\dagger}
\left(
\begin{matrix}
0 & 1 & 0 \\
0 & 0 & -1 \\
-1 & 0 & 0
\end{matrix}\right)
{\bf a}_i.
\end{align}
Hopping from the left to the right therefore promotes a particle
counterclockwise along the triangle of Fig. \ref{Fig1}b: Upon hopping to the
right, a $\ket{+}$ particle becomes $\ket{0}$, a $\ket{0}$ becomes $\ket{-}$,
and a $\ket{-}$ becomes $\ket{+}$. This makes it straightforward to write down
the hopping in terms of raising/lowering operators, and one obtains a relatively
compact second-order Hamiltonian:
\begin{widetext}
\begin{align}
 h_{x}=& -\frac{t^2}{U} \Big\{
\lambda_i^{(4)} \lambda_{i+1}^{(6)} -
\lambda_i^{(5)} \lambda_{i+1}^{(7)} -
\lambda_i^{(1)} \lambda_{i+1}^{(4)} +
\lambda_i^{(2)} \lambda_{i+1}^{(5)} -
\lambda_i^{(6)} \lambda_{i+1}^{(1)} -
\lambda_i^{(7)} \lambda_{i+1}^{(2)}
 \nonumber \\ &
-\frac{1}{2} \left[
\lambda_i^{(3)} \lambda_{i+1}^{(3)} +
\lambda_i^{(8)} \lambda_{i+1}^{(8)} +
\sqrt{3}\left(
\lambda_i^{(3)} \lambda_{i+1}^{(8)} -
\lambda_i^{(8)} \lambda_{i+1}^{(3)}
\right)
\right]
\Big\}.
\end{align}
\end{widetext}
As in the previous section, the simulated annealing algorithm has been used to
grasp the spin textures in the ground state of this Hamiltonian. First, we freeze the hopping
in the $y$-direction, and study $h_x$ alone. The spins
now avoid the $\lambda^{(3)}-\lambda^{(8)}$ plane due to the spin-orbit
coupling. In the remaining planes, a three-periodic pattern is exhibited, see
Fig. \ref{Fig4}(a). Different from the case of diagonal spin-orbit coupling, in
none of the planes does the system exhibit ferromagnetic alignment.

If, additionally to the spin-orbit coupling in $x$-direction, one also assumes the diagonal
spin-orbit term along the $y$-direction, according to Eq. (\ref{SOC}), the two terms together
gives rise to periodic spin patterns in both direction. As shown in Fig.
\ref{Fig4}(b), the unit cell of the spin lattice then becomes two-dimensional,
containing $3 \times 3$ sites. 
These ground
state patterns have been consistently obtained for various system sizes
under open boundary conditions, as well as for periodic boundary
conditions
and linear dimensions divisible by 3. Incommensurate textures were obtaind in
the case of periodic boundaries
and system sizes not being divisible by 3. We therefore expect the 3x3 unit
cell to survive in the thermodynamic limit.

\section{Experimental realization}
In this section, we comment on the realization of the scenarios discussed in
this paper, and discuss the different requirements.

\paragraph{Three-component bosons.}
There is a variety of bosonic atoms providing three or more internal spin
degrees of freedom \cite{ueda-review}. In particular, cold atom experiments
frequently work with $^{87}$Rb in the $F=1$ hyperfine manifold, cf.
Ref.\onlinecite{spielman-peierls} for an experiment in an optical lattice. Such
species allow for the realization of SU(3) Bose-Hubbard like systems.

\paragraph{Atomic Mott phase.}
Next, the experimental setup has to realize the strongly interacting limit,
where the particle fluctuations are strongly suppressed as described
theoretically e.g.
in Refs.~\onlinecite{imambekov,eckert,mlbook,sanpera1,sanpera2}. In this limit,
superexchange interactions may lift the huge degeneracy between different spin
states, but the corresponding energy scale, given by $t^2/U$, is necessarily
very small, since $t \ll U$. The stability of the spin textures therefore
requires an immense cooling effort. Advances in cooling technologies have,
during the past years, allowed for experimentally demonstrating superexchange
interactions \cite{Porto-se1,Porto-se2,Bloch-se1,Bloch-se2,leticia1,leticia2}.
In a recent experiment with fermions, the antiferromagnetic correlations due to
the superexchange have been observed \cite{hulet14}.

\paragraph{Spin-orbit coupling.} In recent years, a variety of
proposals for engineering spin-orbit coupling in ultra-cold atoms has been
made, see for
instance Refs.~\onlinecite{dalibard,dalibard-gerbier,mazza1,mazza2}. The first
experimental realization of artificial spin-orbit coupling was achieved in Ref.
\onlinecite{spielman-sobec} using Raman coupling within a pseudospin-1/2
manifold of $F=1$ $^{87}$Rb atoms. In the experiment, a quadratic Zeeman shift
detunes the third state from resonance in order to mimic SU(2), whereas an SU(3)
spin-orbit coupling would arise naturally without this detuning. A theoretical
proposal which explicitly considers spin-1 spin-orbit coupling within the very
flexible $N$-pod scheme is found in Ref. \onlinecite{gediminas-spin1}.

While such flexibility might be needed to engineer the off-diagonal spin-orbit
coupling of Sec. IV, much simpler schemes are sufficient to obtain a diagonal
spin-orbit coupling, that is, a spin-dependent hopping phase. The
implementation of a diagonal SU(2) spin-orbit coupling has been proposed
recently in Ref. \onlinecite{ketterle-soc}, stressing its great advantages by
avoiding the heating related to spin-flip processes. The proposal is based on a
lattice tilt to replace normal hopping by laser-assisted Raman tunneling. This
technique has already given rise to artificial magnetic fields for atoms
in optical lattices \cite{ketterle-harper,aidelsburger13}. In contrast to an
artifificial magnetic field, the artificial spin-orbit coupling does not require
spatially dependent hopping phases, but instead, a spin dependence of the
hopping phases is needed. The latter can be achieved via a
lattice tilt induced by a magnetic field gradient, and the use of hyperfine
states with different magnetic moments. In our case, two spin states experience
the same hopping phase, while the third state does not gain a hopping phase.
Accordingly, two hyperfine states should have the same magnetic moment,
sufficiently different from the magnetic moment of the third state.

\paragraph{Detection of the spin textures.}
Once the desired Hamiltonian has been engineered at sufficiently low
temperature, the main experimental concern is how to detect the ground state
and its properties. The observable related to the spin textures reported in
this paper are the spin-spin correlations, see Eqs. (\ref{qcorr}) and
(\ref{ccorr}). Correlation measurements in a Bose gas have been pioneered using
noise interferometry in Ref. \onlinecite{foelling05}. Nowadays, it is also
possible to address single atoms in a spin-resolved way, using a quantum gas
microscope \cite{greiner-microscope}. Information about the periodicity of the
spin pattern is encoded in the magnetic structure factor, which is directly
measurable with a spin-dependent Bragg scattering \cite{corcovilos}. This
technique makes use of the fact that the light scattered from an atom depends on
the atom's internal state, rendering the interference pattern in a Bragg
scattering experiment spin dependent with constructive interferences along
angles corresponding to peaks in the magnetic structure factor. This technique
has recently been used to observe antiferromagnetic correlations of a quantum
gas in the superexchange limit of the Fermi-Hubbard model \cite{hulet14}.

\section{Summary and Outlook}
We have derived the effective Hamiltonian of three-component bosons corresponding to the Mott phase in the presence of
spin-orbit coupling. We have focused on two simple cases: a purely
diagonal hopping with spin-dependent phases, as well as a particular purely
off-diagonal spin-orbit coupling. The abelian spin-orbit coupling allows us to
introduce an SU(3) analog of the Dzyaloshinskii-Moriya interaction
(antisymmetric in the pseudo-spin positions). We studied the one-dimensional
chain and the square lattice.  Both, the diagonal and the off-diagonal coupling,
lead to SU(3) spin spiral textures in the ground state. The periodicities of the
spin textures are controlled by the spin-orbit coupling parameters. 
For diagonal spin-orbit coupling, remaining SU(3) degeneracies are lifted by
asymmetry between the inter- and intra-particle interactions, $\mu < 1$.
Alternatively, the spiral patterns are obtained by restricting to a Hilbert
space with fixed spin polarization.

On the quantum level, the spin textures can be traced back to a local gauge
transformation over Dicke-like states. The classical ground state phase on the
two-dimensional lattice was studied using the simulated annealing technique,
yielding, for diagonal spin-orbit coupling, phases that can be understood from
the corresponding one-dimensional case. On the other hand, a generic unique
classical ground state spiral phase was obtained for symmetric interactions with
non-abelian spin-orbit coupling. 
 
More generally, the interplay between  diagonal and off-diagonal spin-orbit
coupling  leads to a much richer  landscape of spin Hamiltonians than discussed
here. The work considered here could be extended to the study of the 
competition between the asymmetry of interactions $\mu$ and the various types of
spin-orbit coupling.

A completely different physical scenario might arise in the
two-dimensional case, if one changes the lattice geometry. Assume, for instance,
a triangular lattice with spin-dependent hopping phases along one direction, and
trivial hopping on the other bonds: It is clear that the spiral order favored by
the spin-dependent hopping cannot be implemented without frustrating at least
one of the ferromagnetic bonds along the other direction. This results in a
highly complex quantum problem for which the predictive power of numerical
simulations is restricted, and which might best be addressed by a quantum
simulation.

\begin{acknowledgements}
We acknowledge support from ERC advanced Grant OSYRIS, EU IP SIQS, EU STREP EQuaM (FP7/2007-2013 Grant No.
323714), FOQUS, and Fundaci{\'o} Cellex. R.W.C. acknowledges a Mobility Plus fellowship from the Polish Ministry of Science and Higher
Education and the (Polish) National Science Center Grant No DEC-2011/03/B/ST2/01903. 
\end{acknowledgements}

\appendix
\begin{widetext}
\section{Effective Hamiltonians}
\subsection{SU(3) notation of biquadratic Heisenberg Hamiltonian}
The SU(2) spin-1 matrices read
\begin{align}
 S^x = \frac{1}{\sqrt{2}}
\left(
\begin{matrix}
 0 & 1 & 0 \\
 1 & 0 & 1 \\
 0 & 1 & 0
\end{matrix}
\right),
\ \ \
 S^y = \frac{1}{\sqrt{2}}
\left(
\begin{matrix}
 0 & -i & 0 \\
 i & 0 & -i \\
 0 & i & 0
\end{matrix}
\right),
\ \ \
 S^z =
\left(
\begin{matrix}
 1 & 0 & 0 \\
 0 & 0 & 0 \\
 0 & 0 & -1
\end{matrix}
\right).
\end{align}
The spin-1 vector is thus written in terms of Gell-Mann matrices as
\begin{align}{\bf S} = \frac{1}{\sqrt{2}}
\left(\lambda^{(1)}+\lambda^{(6)},\lambda^{(2)}+\lambda^{(7)}, \frac{1}{\sqrt{2}}(\lambda^{(3)}+\sqrt{3}\lambda^{(8)})\right).
\end{align}
 Using this equation, it is straightforward to translate any spin-1 Hamiltonian
into the SU(3) picture. In particular, for the biquadratic Heisenberg
Hamiltonian, Eq. (\ref{biqua}), we find the following contributions
\begin{align}
 {\bf S}_i \cdot {\bf S}_j =&
\frac{1}{2} \left( \lambda^{(1)}_i\cdot\lambda^{(1)}_j+\lambda^{(2)}
_i\cdot\lambda^{(2)}_j+\lambda^{(6)}_i\cdot\lambda^{(6)}_j+\lambda^{(7)}
_i\cdot\lambda^{(7)}_j \right)   \nonumber \\ &
 +\frac{1}{2} \left(
\lambda^{(1)}_i\cdot\lambda^{(6)}_j+\lambda^{(6)}
_i\cdot\lambda^{(1)}_j+\lambda^{(2)} _i\cdot\lambda^{(7)}_j+\lambda^{(7)}
_i\cdot\lambda^{(2)}_j \right)  \nonumber \\ &
+ \frac{1}{4} \left(
\lambda^{(3)}_i\cdot\lambda^{(3)}_j+3\lambda^{(8)}
_i\cdot\lambda^{(8)}_j+ \sqrt{3} \lambda^{(3)} _i\cdot\lambda^{(8)}_j + \sqrt{3}
\lambda^{(8)}_i\cdot\lambda^{(2)}_j \right),
 \\
\left({\bf S}_i \cdot {\bf S}_j\right)^2  =&  -\frac{1}{2}
\left( \lambda^{(1)}
_i\cdot\lambda^{(6)}_j+\lambda^{(6)}
_i\cdot\lambda^{(1)}_j+\lambda^{(2)}_i\cdot\lambda^{(7)}_j+\lambda^{(7)}
_i\cdot\lambda^{(2)}_j \right) \nonumber \\ &
+ \frac{1}{4} \left( \lambda^{(3)}_i\cdot\lambda^{(3)}_j-\lambda^{(8)}
_i\cdot\lambda^{(8)}_j- \sqrt{3} \lambda^{(3)} _i\cdot\lambda^{(8)}_j- \sqrt{3}
\lambda^{(8)}_i\cdot\lambda^{(2)}_j \right) \nonumber \\ &
+\frac{1}{2} \left( \lambda^{(4)}_i\cdot\lambda^{(4)}_j+\lambda^{(5)}
_i\cdot\lambda^{(5)}_j \right).
\end{align}
Apparently, this sum of both contributions yields the SU(3) Heisenberg
Hamiltonian.

\subsection{Pseudospin interactions - XYZ-type Hamiltonian \label{app-pseudo}}
For pseudospin interactions as given in Eq. (\ref{pseudo}), one obtains the
effective Mott Hamiltonian of Eq. (\ref{effpseudo}), with the parameters:
\begin{align}
 J_1 = J_2 &= \frac{t^2}{2U_{+0}}, \\
 J_4 = J_5 &= \frac{t^2}{2U_{+-}}, \\
 J_6 = J_7 &= \frac{t^2}{2U_{0-}}, \\
 J_3 &= \frac{t^2}{2} \left(\frac{1}{U_{++}} + \frac{1}{U_{00}} -
\frac{1}{U_{+0}} \right), \\
 J_8 &= \frac{t^2}{6} \left(\frac{1}{U_{++}} + \frac{1}{U_{00}}  +
\frac{4}{U_{--}} + \frac{1}{U_{+0}} - \frac{2}{U_{+-}} - \frac{2}{U_{0-}}
\right), \\
 J_{38} &= \frac{\sqrt{3}t^2}{6} \left( \frac{1}{U_{++}}-
\frac{1}{U_{--}} + \frac{1}{U_{0-}} - \frac{1}{U_{+-}} \right), \\
h_3 &= \frac{1}{3}\left(\frac{1}{U_{++}} - \frac{1}{U_{00}} \right), \\
h_8 &= \frac{\sqrt{3}t^2}{18} \left(\frac{2}{U_{++}} + \frac{2}{U_{00}}  -
\frac{4}{U_{--}} + \frac{2}{U_{+0}} - \frac{1}{U_{+-}} -
\frac{1}{U_{0-}} \right).
\end{align}

\subsection{Spinor interaction and tunneling phases}
Assuming spinor interactions, Eq. (\ref{spinor-int}), in combination with a
tunneling phase $\alpha$ for particles in $\ket{\pm}$, the following
second-order Hamiltonian is obtained along the direction of tunneling:
\begin{align}
 H_{\rm eff} =& \frac{-t^2}{U_0+U_2} \sum_i \Bigg\{ \Big[  \cos\alpha \left(
\lambda^{(1)}_i\cdot\lambda^{(1)}_{i+1}+\lambda^{(2)}
_i\cdot\lambda^{(2)}_{i+1}+\lambda^{(6)}_i\cdot\lambda^{(6)}_{i+1}+\lambda^{(7)}
_i\cdot\lambda^{(7)}_{i+1} \right)
\\ &
+ \sin\alpha \left(
\lambda^{(1)}_i\cdot\lambda^{(2)}_{i+1}-\lambda^{(2)}
_i\cdot\lambda^{(1)}_{i+1}+\lambda^{(6)}_i\cdot\lambda^{(7)}_{i+1}-\lambda^{(7)}
_i\cdot\lambda^{(6)}_{i+1} \right)  + \frac{4}{3}\lambda_i^{(8)} \cdot
\lambda_{i+1}^{(8)}
 \nonumber \\ &
+ \frac{2\sqrt{3}}{3} \left(
\lambda^{(3)}_i\cdot\lambda^{(8)}_{i+1}+\lambda^{(8)}
_i\cdot\lambda^{(3)}_{i+1} \right) + \frac{2}{3} \lambda_i^{(3)}
-\frac{2\sqrt{3}}{9} \lambda_i^{(8)} \Big]
 \nonumber \\ &
-\frac{3t^2}{3U_0-U_2(1-4\cos\alpha)} \Big[ \frac{2}{3}
 \left(
\lambda^{(1)}_i\cdot\lambda^{(6)}_{i+1}+\lambda^{(6)}
_i\cdot\lambda^{(1)}_{i+1}+\lambda^{(2)}_i\cdot\lambda^{(7)}_{i+1}+\lambda^{(7)}
_i\cdot\lambda^{(2)}_{i+1} \right)
\nonumber \\ &
+ \frac{1}{3} \left(
 \lambda^{(4)}_i\cdot\lambda^{(4)}_{i+1}+\lambda^{(5)}
_i\cdot\lambda^{(5)}_{i+1} + 2
\lambda^{(3)}_i\cdot\lambda^{(3)}_{i+1}
\right)
\nonumber \\ &
- \frac{\sqrt{3}}{3} \left(
\lambda^{(3)}_i\cdot\lambda^{(8)}_{i+1}+\lambda^{(8)}
_i\cdot\lambda^{(3)}_{i+1} \right) - \frac{2}{3} \lambda_i^{(3)}
+\frac{2\sqrt{3}}{9} \lambda_i^{(8)}
\Big]
\nonumber \\ &
-\frac{3t^2}{3U_0-2 U_2(1+2\cos\alpha)} \Big[
 \frac{2}{3} \left(
 \lambda^{(4)}_i\cdot\lambda^{(4)}_{i+1}+\lambda^{(5)}
_i\cdot\lambda^{(5)}_{i+1}
\right)
\nonumber \\ &
- \frac{2}{3}
 \left(
\lambda^{(1)}_i\cdot\lambda^{(6)}_{i+1}+\lambda^{(6)}
_i\cdot\lambda^{(1)}_{i+1}+\lambda^{(2)}_i\cdot\lambda^{(7)}_{i+1}+\lambda^{(7)}
_i\cdot\lambda^{(2)}_{i+1} \right)
\nonumber \\ &
- \frac{\sqrt{3}}{3} \left(
\lambda^{(3)}_i\cdot\lambda^{(8)}_{i+1}+\lambda^{(8)}
_i\cdot\lambda^{(3)}_{i+1} \right) + \frac{1}{3}
\left(
 \lambda_i^{(3)} \cdot \lambda_{i+1}^{(3)} -
 \lambda_i^{(8)} \cdot \lambda_{i+1}^{(8)}
\right)
\Big]
\Bigg\}.
\end{align}
For non-zero tunneling phases, a SU(3)-field is acting in the $\lambda^{(3)}-\lambda^{(8)}$
plane, favoring the polarization of the system within
this plane.
\end{widetext}

\end{document}